\documentclass[
]{ceurart}

\usepackage{listings}
\usepackage{caption}\usepackage{wasysym}

\usepackage{adjustbox}
\definecolor{codegreen}{rgb}{0,0.6,0}
\definecolor{codegray}{rgb}{0.5,0.5,0.5}
\definecolor{codepurple}{rgb}{0.58,0,0.82}
\definecolor{backcolour}{rgb}{0.95,0.95,0.92}

\lstdefinestyle{mystyle}{
    backgroundcolor=\color{backcolour},   
    commentstyle=\color{codegreen},
    keywordstyle=\color{magenta},
    numberstyle=\tiny\color{codegray},
    stringstyle=\color{codepurple},
    basicstyle=\ttfamily\footnotesize,
    breakatwhitespace=false,         
    breaklines=true,                 
    captionpos=b,                    
    keepspaces=true,                 
    numbers=left,                    
    numbersep=5pt,                  
    showspaces=false,                
    showstringspaces=false,
    showtabs=false,                  
    tabsize=2
}

\lstnewenvironment{tabularlstlisting}[1][]
 {%
  \lstset{aboveskip=-1.3ex,belowskip=-3.5ex,#1}%
 }
 {}

\usepackage{pifont}
\begin{document}

\copyrightyear{2022}
\copyrightclause{Copyright for this paper by its authors.
  Use permitted under Creative Commons License Attribution 4.0
  International (CC BY 4.0).}

\conference{--}

\title{Vanilla-Converter: A Tool for Converting Camunda 7 BPMN Models into Camunda 8 Models}

\author[1]{Dragana Sunaric}[%
orcid=0009-0004-0115-0817,
email=dsunaric@gmx.at,
]
\author[1]{Charlotte Verbruggen}[%
orcid=0000-0003-0418-2633,
email=charlotte.verbruggen@tuwien.ac.at,
url=https://informatics.tuwien.ac.at/people/charlotte-verbruggen,
]
\author[1]{Dominik Bork}[%
orcid=0000-0001-8259-2297,
email=dominik.bork@tuwien.ac.at,
url=https://model-engineering.info/,
]

\address[1]{TU Wien, Business Informatics Group,
  Erzherzog-Johann-Platz 1, 1040 Vienna, Austria}

\begin{abstract}
As organizations prepare for the end-of-life of Camunda 7, manual migration remains complex due to fundamental differences between the two platforms. 
We present Vanilla-Converter, a command-line tool that facilitates the migration of BPMN models from Camunda 7 to Camunda 8. 
Vanilla-Converter automates the transformation process, supports a wide range of BPMN elements, and produces a transformed model and a detailed transformation log indicating automatic changes and remaining manual conversion tasks. 
The tool's effectiveness is demonstrated through three case studies with real industrially used Camunda 7 models, confirming its ability to convert these models into valid and executable Camunda 8 models.
\end{abstract}

\begin{keywords}
  BPMN \sep
  Camunda 7 \sep
  Camunda 8 \sep
  Workflow Automation \sep
  Open-source \sep
  VanillaBP \sep
  Model transformation
\end{keywords}

\maketitle

\section{Introduction}
In the field of business process automation, BPMN has established itself as the de facto standard~\cite{CHINOSI2012124,bpmn20}. Enterprises benefit from using a tool like Camunda~\cite{camunda} to execute and control processes modeled using BPMN~\cite{10171706}. Although BPMN is widely accepted, tool-specific variants and extensions of the standard are common~\cite{zarour2020bpmnExtensions,10.1007/978-3-662-45501-2_4}. This is particularly relevant in the context of the announced end-of-life for Camunda 7 in 2030~\cite{c7-eol}, which is driving many organizations to migrate to Camunda 8.

The migration from Camunda 7 to Camunda 8 presents a significant challenge due to the fundamental architectural and operational differences between the two platforms. Camunda 7 is a monolithic, thread-based workflow engine typically embedded within Java applications, whereas Camunda 8 is a cloud-native, distributed system built on Zeebe, relying on external workers and event-driven execution semantics~\cite{camunda-differences}. While some open-source initiatives, such as VanillaBP~\cite{vanillabp-website}, attempt to promote engine-agnostic BPMN modeling through architectural patterns like hexagonal architecture, tooling specifically focused on automating the migration from Camunda 7 to Camunda 8 remains scarce. A further complication arises from the fact that, despite both versions relying on the BPMN standard for process definitions, each extends the standard with incompatible, engine-specific elements. As a result, direct reuse of Camunda 7 models in Camunda 8 is not possible, forcing developers to manually reconstruct models during migration.

Building on the goals of the VanillaBP~\cite{vanillabp-website} initiative, this paper presents a command-line tool designed to bridge the gap between Camunda 7 and Camunda 8 by automating the transformation of BPMN models. The tool facilitates a more efficient and less error-prone migration process, reducing manual effort and improving model consistency across platforms.

\section{Tool Overview}
\begin{figure}[t]
    \centering
    \includegraphics[width=\textwidth]{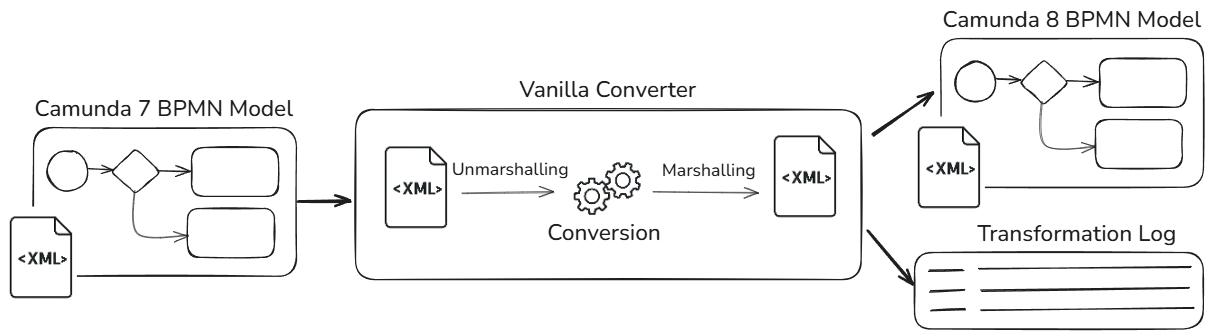}
    \caption{Overview of the Vanilla-Converter}
    \label{fig:overview}
\end{figure}

\textbf{Vanilla-Converter} is a command-line application designed to transform Camunda 7 BPMN into models that are compatible with Camunda 8. It processes Camunda 7 \texttt{.bpmn} files, applies a set of transformation rules, and outputs updated models suitable for use with Camunda 8. Alongside the converted model, the tool generates a detailed transformation log that documents which elements were automatically converted and highlights remaining manual tasks for elements that require user intervention. Fig.~\ref{fig:overview} shows an overview of the tool.

\subsection{Usage} 
The Vanilla-Converter is operated through a simple command-line interface. Users can convert either a single BPMN file or batch-convert all BPMN files within a specified directory:
\begin{itemize}
\item To transform a single BPMN file:
\begin{lstlisting}
java -jar vanilla-transformer.jar /path/to/your/file.bpmn
\end{lstlisting}
\item To transform all BPMN files in a directory:
\begin{lstlisting}
java -jar vanilla-transformer.jar /path/to/your/files
\end{lstlisting}
\end{itemize}
The tool generates transformed BPMN files by appending the suffix \texttt{-transformed} to the original filenames and saving them in the same directory as the input files. During execution, a detailed transformation log is produced, documenting both the automatically applied changes and any remaining manual tasks. This log is output to the console and persisted in the file \textit{logs/transformation.log} for further inspection.

\subsection{Transformation}
The Vanilla-Converter accepts a Camunda 7 \texttt{.bpmn} file as input and processes it directly at the XML level. To enable structured access and manipulation of the BPMN content, the XML file is unmarshalled into a corresponding Java object graph using JAXB (Java Architecture for XML Binding). Although both Camunda 7 and Camunda 8 offer model APIs for working with BPMN files, these APIs are incompatible and do not support cross-version transformations. As a result, our tool avoids using either API and instead performs all manipulations directly on JAXB-generated classes that reflect the XML structure of the BPMN model.
One aspect of this project is the development of custom XML Schema Definitions (XSDs) for both the Camunda 7 and Camunda 8 BPMN specifications. As official XML schemas are not publicly provided by Camunda, these XSDs were manually constructed based on available documentation and the structure of BPMN files generated by each respective version. The resulting schemas serve as a foundation for validating and processing models during transformation. They are available in the project’s GitHub repository~\cite{schemas}.
The transformation logic itself is implemented as a depth-first traversal of the BPMN XML tree. During this traversal, each Camunda 7 element is systematically mapped to its corresponding Camunda 8 representation. Table~\ref{table:mapping} shows an excerpt from the Camunda 7 to 8 mapping. A comprehensive mapping table documenting the supported element conversions is provided in the project’s GitHub repository~\cite{vanilla-converter}.

Once the transformation is complete, the modified Java object graph is marshalled back into a valid Camunda 8 BPMN file using JAXB. The resulting model is syntactically well-formed and fully compatible with the Camunda 8 engine.

\begin{table}[]
\caption{Excerpt from the Camunda 7 to 8 mapping table}
\label{table:mapping}
\scriptsize
\lstset{
  basicstyle=\scriptsize\ttfamily,
  numbers=left,
  numberstyle=\tiny,
  stepnumber=1,
  numbersep=5pt,              
  xleftmargin=1.5em,          
  framexleftmargin=1.5em      
}
\begin{tabular}{|m{2cm}|m{6cm}|m{6cm}|}
\hline
\textbf{BPMN Element} & \textbf{Camunda 7 XML Representation} & \textbf{Mapped to Camunda 8 Element} \\ \hline
Service Task using Delegate Expression & 
\begin{lstlisting}
<bpmn:serviceTask id="Service-Task-2" 
name="DelegateExpression" 
camunda:delegateExpression="${SomeDelegateExpression}">
</bpmn:serviceTask>
\end{lstlisting} 
& 
\begin{lstlisting}
<bpmn:serviceTask name="DelegateExpression" id="Service-Task-2">
    <bpmn:extensionElements>
        <zeebe:taskDefinitiontype="SomeDelegateExpression"/>
    </bpmn:extensionElements>
</bpmn:serviceTask>
\end{lstlisting} 
\\ \hline

Sequence Flow with a condition Expression (mapping to FEEL expression) & 
\begin{lstlisting}
<bpmn:sequenceFlow id="Flow_1"sourceRef="src"targetRef="target">
    <bpmn:conditionExpression xsi:type="bpmn:tFormalExpression">
        ${false}
    </bpmn:conditionExpression>
</bpmn:sequenceFlow>
\end{lstlisting} 
& 
\begin{lstlisting}
<bpmn:sequenceFlow id="Flow_1"sourceRef="src"targetRef="target">
    <bpmn:conditionExpression xsi:type="bpmn:tFormalExpression">
        =false
    </bpmn:conditionExpression>
</bpmn:sequenceFlow>
\end{lstlisting} 
\\ \hline

Sequencial Multi Instance (embedded in Task) & 
\begin{lstlisting}
<bpmn:multiInstanceLoopCharacteristics 
camunda:asyncBefore="true" 
camunda:collection="${collection}" 
camunda:elementVariable="${element}"  
isSequential="true" />
\end{lstlisting} 
& 
\begin{lstlisting}
<bpmn:multiInstanceLoopCharacteristics isSequential=" true">    
    <bpmn:extensionElements>
        <zeebe:loopCharacteristics inputCollection="= collection" inputElement="element" />
    </bpmn:extensionElements>
</bpmn:multiInstanceLoopCharacteristics>
\end{lstlisting} 
\\ \hline
\end{tabular}
\end{table}

\subsection{Transformation Log}
As part of the transformation process, Vanilla-Converter generates a detailed transformation log that provides transparency and traceability for each applied transformation. This log documents both the automated transformations performed by the tool and any remaining manual tasks that must be addressed by the user. These manual tasks are necessary not only to achieve syntactic compatibility with Camunda 8, but also to ensure the resulting process model is fully executable and semantically valid within a real-world application context.

The transformation proceeds hierarchically, traversing the BPMN model from outer to inner elements, similar to a depth-first traversal of a tree. The log reflects this traversal structure, making it easier for users to understand the transformation context of nested elements.

The log output uses the following standardized messages:
\begin{itemize}
\item \textit{MAPPING} indicates that the transformation of a specific BPMN element has begun.
\item \textit{FINISHED MAPPING:} marks the successful completion of a mapping for the element.
\item \textit{NO MAPPING NEEDED:} denotes elements that are structurally and semantically compatible between Camunda 7 and Camunda 8 and therefore do not require transformation.
\item \textit{TODO} flags elements where manual intervention is required to complete the migration. These tasks are typically essential for the correct execution of the model in Camunda 8.
\item \textit{TODO (OPTIONAL)} highlights optional tasks where user action may improve the model or resolve potential ambiguities but may not be strictly required.
\end{itemize}
An excerpt from a representative transformation log is shown below:

\begin{lstlisting}
MAPPING: bpmn:event with id=Event_0j8p
MAPPING: bpmn:throwEvent with id=Event_0j8p
MAPPING: bpmn:eventDefinition with id=MessageEventDefinition_1nr2ae9
TODO: manually configure Jobworker for Message Event Definition with id=MessageEventDefinition_1nr2ae9
FINISHED MAPPING: bpmn:eventDefinition with id=MessageEventDefinition_1nr2ae9
MAPPING: bpmn:delegateExpression configureJobType into zeebe:taskDefinition
MAPPING: camunda:expression configureJobType into FEEL Expression Language
TODO (OPTIONAL): set retries=?? in zeebe:taskDefinition Element
TODO (OPTIONAL): adapt zeebe:taskDefinition type for Event with id=Event_0j8p to select correct JobWorker
FINISHED MAPPING: bpmn:throwEvent with id=Event_0j8p
FINISHED MAPPING: bpmn:event with id=Event_0j8p
\end{lstlisting}
This structured output enables users to quickly identify which parts of the model have been successfully transformed and which require manual completion, thereby supporting a more reliable and efficient migration process.

\subsection{Supported Elements} 
The tool supports a broad range of BPMN elements commonly used in real-world business process models. Elements that are functionally equivalent in both Camunda 7 and Camunda 8 are transferred without modification. In cases where elements have been deprecated or are no longer supported in Camunda 8—such as Conditional Events or Cancel Events—the tool detects them during the transformation and logs appropriate warnings in the transformation log. Currently, the only Camunda 7 element that remains unsupported by the tool is the Script Task. Support for this element is planned as part of future development.

For each BPMN element encountered, the tool provides detailed log output indicating whether a mapping was successfully performed, whether manual intervention is required, or whether the element itself is entirely incompatible with Camunda 8. This transparency allows users to easily identify and address remaining transformation gaps before deploying the resulting model.

\section{Maturity of the Tool}
The maturity of Vanilla-Converter was evaluated through three case studies.
The first case study involved the TaxiRide Blueprint Project~\cite{taxiride-project}, a publicly available reference implementation that exists in both Camunda 7 and Camunda 8 versions. The converter was applied to the Camunda 7 version of the model, and the output was qualitatively compared against the official Camunda 8 implementation. The transformation was completed successfully, with the tool automatically converting the core process structure. The transformation log highlighted ten minor manual adaptation tasks, such as defining correlation keys for message events, which are required to make the model fully executable on Camunda 8. This demonstrates that the tool is capable of handling real-world models with high fidelity, while still offering clear guidance on necessary post-processing steps. 
The other two case studies involved a quantitative evaluation using the \textit{ELMO} project~\cite{elmo-project} and Camunda Platform Examples~\cite{camunda7-exmaple}. Further details of both evaluation cases are provided in the project's GitHub repository~\cite{case-studies}. In all case studies, the tool successfully transformed the Camunda 7 models into valid Camunda 8 models, including ten process models with a combined total of 229 BPMN elements in the TaxiRide project, one process model with 33 elements in the ELMO project, and 39 processes with 413 elements in the Camunda Platform Examples. The resulting models were executable in the Camunda 8 engine without BPMN validation errors or warnings, confirming the structural correctness and semantic soundness of the transformation. Additionally, the tool generated a transformation log detailing all applied mappings and highlighting remaining manual conversion tasks (TODOs) required to make the processes fully runnable.

\subsection{Open Topics and Future Work}
The current implementation of the converter is limited to unidirectional transformations from Camunda 7 to Camunda 8. Supporting reverse transformations (i.e., from Camunda 8 back to Camunda 7) or even enabling interoperability between other BPMN-based workflow engines such as Activiti remains an open challenge. A key obstacle is the fact that different workflow engines often interpret and extend the BPMN standard in proprietary ways, resulting in compatibility issues during model transformation.
A promising direction for future research lies in conducting a systematic analysis and comparison of BPMN extensions across major process engines. Such an effort could inform the development of a unified extension framework or modeling profile that accommodates engine-specific requirements while preserving core BPMN semantics. This would promote greater interoperability and foster a more engine-agnostic approach to BPMN-based process modeling.

\section{Conclusion}
In this paper, we introduced Vanilla-Converter, an open-source command-line tool for automatically converting BPMN models in Camunda 7 format into equivalent Camunda 8 models. 
The tool shall ease the transition to the latest Camunda version. 
The source code for the Vanilla-Converter and all details on the mapping between the two Camunda versions, as well as the case studies performed, are available at \url{ https://github.com/dsunaric/vanilla-converter}. 
A demo video showing how the tool works is available at \url{https://youtu.be/tLOUjrNbE0w}.

\bibliography{sample-ceur}

\begin{thebibliography}{15}
\expandafter\ifx\csname natexlab\endcsname\relax\def\natexlab#1{#1}\fi
\providecommand{\url}[1]{\texttt{#1}}
\providecommand{\href}[2]{#2}
\providecommand{\path}[1]{#1}
\providecommand{\DOIprefix}{doi:}
\providecommand{\ArXivprefix}{arXiv:}
\providecommand{\URLprefix}{URL: }
\providecommand{\Pubmedprefix}{pmid:}
\providecommand{\doi}[1]{\href{http://dx.doi.org/#1}{\path{#1}}}
\providecommand{\Pubmed}[1]{\href{pmid:#1}{\path{#1}}}
\providecommand{\bibinfo}[2]{#2}
\ifx\xfnm\relax \def\xfnm[#1]{\unskip,\space#1}\fi
\bibitem[{Chinosi and Trombetta(2012)}]{CHINOSI2012124}
\bibinfo{author}{M.~Chinosi}, \bibinfo{author}{A.~Trombetta},
\newblock \bibinfo{title}{Bpmn: An introduction to the standard},
\newblock \bibinfo{journal}{Computer Standards \& Interfaces} \bibinfo{volume}{34} (\bibinfo{year}{2012}) \bibinfo{pages}{124--134}. \URLprefix \url{https://www.sciencedirect.com/science/article/pii/S0920548911000766}. \DOIprefix\doi{https://doi.org/10.1016/j.csi.2011.06.002}.
\bibitem[{{Object Management Group}(2011)}]{bpmn20}
\bibinfo{author}{{Object Management Group}}, \bibinfo{title}{Business Process Model and Notation (BPMN) Version 2.0}, \bibinfo{type}{Technical Report} \bibinfo{number}{formal/2011-01-03}, Object Management Group (OMG), \bibinfo{year}{2011}. \URLprefix \url{https://www.omg.org/spec/BPMN/2.0}, \bibinfo{note}{accessed: 2025-05-17}.
\bibitem[{cam(nd)}]{camunda}
\bibinfo{title}{Camunda bpm}, \bibinfo{howpublished}{\url{https://camunda.com/}}, \bibinfo{year}{n.d.} \bibinfo{note}{Accessed: 2025-05-17}.
\bibitem[{David et~al.(2023)David, Zmaranda, Györödi, and Györödi}]{10171706}
\bibinfo{author}{G.~David}, \bibinfo{author}{D.~R. Zmaranda}, \bibinfo{author}{R.-S. Györödi}, \bibinfo{author}{C.~A. Györödi},
\newblock \bibinfo{title}{Exploring the impact of workflow engines on business process management in enterprise applications. a case-study: Camunda},
\newblock in: \bibinfo{booktitle}{2023 17th International Conference on Engineering of Modern Electric Systems (EMES)}, \bibinfo{year}{2023}, pp. \bibinfo{pages}{1--4}. \DOIprefix\doi{10.1109/EMES58375.2023.10171706}.
\bibitem[{Zarour et~al.(2020)Zarour, Benmerzoug, Guermouche, and Drira}]{zarour2020bpmnExtensions}
\bibinfo{author}{K.~Zarour}, \bibinfo{author}{D.~Benmerzoug}, \bibinfo{author}{N.~Guermouche}, \bibinfo{author}{K.~Drira},
\newblock \bibinfo{title}{A systematic literature review on bpmn extensions},
\newblock \bibinfo{journal}{Business Process Management Journal} \bibinfo{volume}{26} (\bibinfo{year}{2020}) \bibinfo{pages}{1473--1503}. \URLprefix \url{https://doi.org/10.1108/BPMJ-01-2019-0040}. \DOIprefix\doi{10.1108/BPMJ-01-2019-0040}.
\bibitem[{Braun and Esswein(2014)}]{10.1007/978-3-662-45501-2_4}
\bibinfo{author}{R.~Braun}, \bibinfo{author}{W.~Esswein},
\newblock \bibinfo{title}{Classification of domain-specific bpmn extensions},
\newblock in: \bibinfo{editor}{U.~Frank}, \bibinfo{editor}{P.~Loucopoulos}, \bibinfo{editor}{{\'O}.~Pastor}, \bibinfo{editor}{I.~Petrounias} (Eds.), \bibinfo{booktitle}{The Practice of Enterprise Modeling}, \bibinfo{publisher}{Springer Berlin Heidelberg}, \bibinfo{address}{Berlin, Heidelberg}, \bibinfo{year}{2014}, pp. \bibinfo{pages}{42--57}.
\bibitem[{c7-(nd)}]{c7-eol}
\bibinfo{title}{Camunda 7 enterprise end of life (eol) extension announcement}, \bibinfo{howpublished}{\url{https://camunda.com/blog/2025/02/camunda-7-enterprise-end-of-life-extension/}}, \bibinfo{year}{n.d.} \bibinfo{note}{Accessed: 2025-05-17}.
\bibitem[{cam(nd)}]{camunda-differences}
\bibinfo{title}{Migrate from camunda 7 - conceptual differences}, \bibinfo{howpublished}{\url{https://docs.camunda.io/docs/guides/migrating-from-camunda-7/conceptual-differences/}}, \bibinfo{year}{n.d.} \bibinfo{note}{Accessed: 2025-05-17}.
\bibitem[{van(nd)}]{vanillabp-website}
\bibinfo{title}{Vanillabp}, \bibinfo{howpublished}{\url{https://www.vanillabp.io/}}, \bibinfo{year}{n.d.} \bibinfo{note}{Accessed: 2025-05-17}.
\bibitem[{sch(nd)}]{schemas}
\bibinfo{title}{Xml schemas for camunda 7 and camunda 8}, \bibinfo{howpublished}{\url{https://github.com/dsunaric/vanilla-converter/tree/main/src/main/resources/schema}}, \bibinfo{year}{n.d.}
\bibitem[{Sunaric(2025)}]{vanilla-converter}
\bibinfo{author}{D.~Sunaric}, \bibinfo{title}{Vanilla converter}, \bibinfo{howpublished}{\url{https://github.com/dsunaric/vanilla-converter}}, \bibinfo{year}{2025}.
\bibitem[{Pelikan(2024{\natexlab{a}})}]{taxiride-project}
\bibinfo{author}{S.~Pelikan}, \bibinfo{title}{Taxi ride blueprint for business-processing (micro)services}, \bibinfo{howpublished}{\url{https://github.com/phactum/taxiride-blueprint}}, \bibinfo{year}{2024}{\natexlab{a}}. \bibinfo{note}{Accessed: 2025-05-17}.
\bibitem[{Pelikan(2024{\natexlab{b}})}]{elmo-project}
\bibinfo{author}{S.~Pelikan}, \bibinfo{title}{Elmo}, \bibinfo{howpublished}{\url{https://github.com/stephanpelikan/elmo}}, \bibinfo{year}{2024}{\natexlab{b}}. \bibinfo{note}{Accessed: 2025-05-17}.
\bibitem[{cam(nd)}]{camunda7-exmaple}
\bibinfo{title}{Camunda platform examples}, \bibinfo{howpublished}{\url{https://github.com/camunda/camunda-bpm-examples}}, \bibinfo{year}{n.d.}
\bibitem[{Sunaric(2025)}]{case-studies}
\bibinfo{author}{D.~Sunaric}, \bibinfo{title}{Vanilla converter - case studies}, \bibinfo{howpublished}{\url{https://github.com/dsunaric/vanilla-converter/tree/main/src/main/resources/examples}}, \bibinfo{year}{2025}.

\end{thebibliography}

\appendix

\end{document}